\documentstyle[aps,epsfig,float,twocolumn,prl]{revtex}
\newcommand{\be}{\begin{equation}}
\newcommand{\ee}{\end{equation}}
\begin{document}
\twocolumn[\hsize\textwidth\columnwidth\hsize\csname @twocolumnfalse\endcsname
\draft
\title{Random Vibrational Networks and Renormalization Group}
\author{M. B. Hastings}
\address{
Center for Nonlinear Studies and Theoretical Division, Los
Alamos National Laboratory, Los Alamos, NM 87545,
hastings@cnls.lanl.gov }
\date{December 10, 2002}
\maketitle
\begin{abstract}
We consider the properties of vibrational dynamics on random networks, with
random masses and spring constants.  The localization properties of the
eigenstates contrast greatly with the Laplacian case on these networks.
We introduce several real-space renormalization techniques which can be
used to describe this dynamics on general networks, drawing on strong disorder
techniques developed for regular lattices.  The renormalization group is
capable of elucidating the localization properties, and provides, even for 
specific network instances, a fast approximation technique for determining the 
spectra which compares well with exact results.
\vskip2mm
\end{abstract}
]
The study of linear equations with random coefficients has a long
history, dating back to the study of random matrices\cite{mehta}, systems
with no spatial structure at all.
Later work incorporated spatial structure,
focusing on the case of {\it regular} lattices,
starting with Anderson's work studying electron localization
via the Schroedinger equation\cite{anderson}.
The Bethe lattice was also introduced
as a soluble case\cite{aat}, illustrating the localization transition in
these systems.

Recently, the study of dynamics on more general {\it networks} has become
of interest\cite{net}.  The systems above can be included in the
case of networks, as can additional systems, such as sparse random
matrices, or small world networks\cite{sw}.
Networks contrast with regular lattices, as the concept of locality on
networks may not be well defined.  This concept of
locality is essential to physics, where
so many subjects rely on introducing slowly
varying ``hydrodynamic"
fields to describe the long wavelength dynamics.  Small world
networks illustrate the problem: there is always a route between any two nodes
in the network which involves traversing sites which are near each
other in physical distance, but there may be a much shorter route which
involves using some long jumps.

In this work, we consider the specific problem of a
random networks of masses connected by springs.  This problem is
chosen for several reasons.  It includes, but is richer than, the related
problem of a Laplacian on a network\cite{diff}.  It also has physical
applications to vibrational modes in amorphous and granular
media\cite{amorph}.  Finally,
this problem is a good problem for studying the application of
real-space renormalization group techniques, originally
developed for regular lattices, to arbitrary networks, including
random graphs and small-world networks.
Real-space techniques were originally developed for application to
statistical mechanics systems without randomness.  Since then, they have been
applied to a number of disordered systems, in some cases yielding exact
results\cite{fisher}.  The advantage of these techniques is that, without
assuming any specific structure of the system as required for momentum
space techniques, they are able to reveal the geometry of the 
network\cite{newman}.
We will see that not only are these techniques able to provide very
accurate approximations, but also to find the correct notion of
locality for the given network.  

{\it Random Networks, Random Masses---}
We consider a random network of masses connected by springs.  This generalizes
the problem studied by Dyson\cite{dyson} in one-dimension.  The equation
of motion for a particle $i$ of mass $m_i$ connected by springs with spring
constants $k_{ij}$ to other particles $j$ is
\be
\partial^2_t x_i=-\sum\limits_j O_{ij} x_j,
\ee
where the matrix $O$ has elements $O_{ij}=-k_{ij}/m_{i}$ for $i\neq j$
and $O_{ii}=\sum\limits_j k_{ij}/m_{i}$ (we define $O$ with these
signs so its eigenvalues will be positive).  By a rescaling of coordinates,
$x_i \rightarrow m_i^{1/2} x_i$, we can study instead the matrix $L$ defined
by $L_{ij}=m_i^{1/2} O_{ij} m_j^{-1/2}$.  Then, $L$ has the same eigenvalues
as $O$, but is a symmetric matrix.  

It is possible to relate this problem to a
random hopping problem, studied in quantum
mechanics.  Introduce a matrix $H_{(jk),i}$, where the index $i$ labels
a site and the index $(jk)$ labels a link between sites (by definition, the
links $(jk)$ and $(kj)$ are the same).  Then, take
$H_{i,(ij)}=H_{(ij),i)}=\sqrt{k_{ij}/m_{i}}$ and 
$H_{j,(ij)}=H_{(ij),j)}=-\sqrt{k_{ij}/m_{j}}$.
The matrix $H$ connects sites to links and vice-versa.
The matrix $H^2$ is block-diagonal: it connects sites to sites and links to 
links.  Restricting to just the block of $H^2$ which connects sites to sites,
we find that $L=H^2$.

Automatically, $O$ has a zero mode
associated with a uniform motion of all the masses.  This will be essential
to the renormalization procedure defined below, as we will study the
behavior of $O$ for small eigenvalues.  
In the case that all $m_i=1$,
$L$ is a Laplacian on the network with
different couplings $k_{ij}$ between nodes.  If instead 
$m_i=1/n_i$,
with $n_i$ the number of nodes connected to $i$, and
$k_{ij}=0,1$ depending on whether nodes $i,j$ are connected, then
we obtain another definition of the network Laplacian\cite{fc}.
We will find that including random masses leads to, in many cases,
greatly different localization properties than without; further, we
will see that the renormalization procedures naturally lead to
variations in the mass.

{\it Renormalization---}
Inspired by real-space techniques applied to strong disorder systems
in one dimension, we consider a variety of renormalization procedures.
The previous work\cite{fisher} was based on a matrix such as $H$.  
Although it is not essential to the discussion below, we
recall that the technique for operator $H$ goes as follows:
pick the largest term in $H$.  Let this term have magnitude $\Lambda$.
This term connects a site $j$ and a link $(jk)$.  
Let site $j$ also be connected to link $(ij)$, while link $(jk)$ is
also connected to site $k$.  The approximation behind the renormalization
procedure consists in assuming that
$\Lambda$ is much larger than all other terms in $H$.  Then,
$H$ has two eigenvalues $\approx \pm \Lambda$.  We then remove site $j$ and
link $(jk)$ from the system and use second order perturbation theory to
connect the link $(ij)$ to $k$.  We find that that the resulting term in
the renormalized $H$ is $H_{(ij),k}=H_{(ij),j}H_{(jk),k)}/H_{j,(jk)}$.

Since $H^2=L$, the renormalization of $H$ will enable us to renormalize $L$.
Based on this fact, and the above procedure, we propose the following
technique that can be applied directly to $L$: (1) choose the site $i$
with the largest $L_{ii}$.  (2) Make the approximation that this $L_{ii}$ is 
much larger than $|L_{ij}|$ for any other $j$, and thus declare that
$L$
has an eigenvalue equal to $L_{ii}$ with eigenvector concentrated
on site $i$.  (3) Remove the site $i$ from the network and define a new
matrix $\tilde L$ connecting the remaining sites with $\tilde L_{jk}=
L_{jk}+L_{ji}L_{ik}/L_{ii}$.  We find that if $i$ has only
the two neighbors, $j$ and $k$, then the masses of sites $j,k$ are
unchanged by this procedure, while the sites are connected by
a spring constant $k_{jk}=k_{ji}k_{ik}/(k_{ji}+k_{jk})$.

This procedure works if $L_{ii}$ is indeed much larger in
than $|L_{ij}|$.
In one dimension, this requires that the mass of site $i$ be much smaller
than that of either of its neighbors.  
If the elements of $H$ are
chosen randomly from a distribution with finite width, this leads
to a random distribution of mass ratios of nearest neighbors in the
original lattice.  Then, since $m_k/m_j$ and $m_j/m_i$ are in this case
uncorrelated, the distribution of the mass ratio $m_k/m_i$ is broader.  Thus,
the distribution of mass ratios broadens under renormalization,
justifying the procedure.

We, however, consider a system in which the masses themselves,
not the mass ratios, are chosen from a
given distribution so that the mass ratios remain narrow under
the renormalization and thus this procedure does not work for a one
dimensional system.  However, if the connectivity $n_i$ of site $i$ is
large, then $L_{ii}/|L_{ij}| \approx n_i>>1$.
Thus for networks, this procedure
can work.  A one dimensional lattice remains a one dimensional lattice
under this procedure, while any other lattice or network changes its
topology\cite{huse}.

Define the Green's function $G_L(E)=1/(E-L)$.  Then, for $E=0$, 
$G_L=G_{\tilde L}$ for the sites that remain in the network, so this
procedure is exact for $E=0$.
For other $E$, it is possible to follow a renormalization procedure with
$\tilde L_{jk}=L_{jk}+L_{ji}L_{ik}/(L_{ii}-E)$\cite{zee}. 

A more powerful renormalization technique is to consider pairs of
sites.  This two-site technique is more accurate at each renormalization step.
It also tends to increase the connectivity of the sites and randomize
the masses, thus leading to a situation in which each renormalization
step is more accurate than the previous.  We proceed as follows:
(1) choose the site $i$ with the largest $L_{ii}$, and then find its
neighbor $j$ with the largest $|L_{ij}|$.  (2) 
Consider the two-by-two submatrix of $L$ which involves only sites $i,j$.
This has eigenvalues $E^{\pm}=(L_{ii}+L_{jj})/2\pm
\sqrt{(L_{ii}-L_{jj})/2+L_{ij}^2}$ with $E^+>E^->0$, and has
eigenvectors $v^{\pm}_{ij}$.
Change basis from $i,j$ to the basis of these eigenvectors $v^{+},v^{-}$ 
to diagonalize this matrix, using
$L_{kv^{\pm}}=L_{ki}v^{\pm}_i+L_{kj}v^{\pm}_j$.
Make the approximation that
$L_{ii}, L_{ij}, L_{jj}$ are much bigger than any other term in
$L$, and thus declare that $L$ has one eigenvalue equal to $E^{+}$, with
corresponding eigenvector $v^{+}$.  
(3) Remove $v^{+}$ from the network and define a new
matrix $\tilde L$ connecting the remaining sites with $\tilde L_{kl}=
L_{kl}+L_{kv^+}L_{v^+l}/E^+$.  

This procedure is the same as
that above, but with a change of basis from $i,j$ to $v^{+},v^{-}$.  
We will see below that this makes the procedure much more accurate.  In
addition to the increased accuracy at each step, the procedure tends
to increase the connectivity: applying this procedure to a one dimensional
lattice produces second-neighbor connections.  The
procedure also leads to variation in the masses: the mass of the
new site $v^{-}$ is {\it not} equal to either $m_i$ or $m_j$.  Thus, the
procedure drives itself to a regime in which it become more accurate, due
to higher connectivity and a broader distribution of masses.  
This procedure shows the correct idea of locality for the network:
we know that, as far as the low-energy dynamics are concerned, $i,j$
are ``close by" in the original network.

{\it Localization Properties---}
We have performed numerical simulations to study localization.
The specific network considered is a random graph of $N$ sites: 
for a given probability $p$, two sites
are connected with that probability.  Then, the spring constant connecting
those sites is assigned from a uniform distribution between $0$ and $1$ (taking
instead all $k_{ij}=0,1$ leads to
little change in the results).  The average
connectivity of a site is equal to $p(N-1)$.
We consider three different mass distributions: (1) $m_i=1$ for all $i$;
(2) $1/\sqrt{m_i}$ is chosen
uniformly between $1$ and $2$, giving
a smooth, bounded distribution of masses; (3)
$1/\sqrt{m_i}$ is chosen uniformly between $0$ and $1$, so that the
distribution of masses is unbounded.

The energy scales with the connectivity of the the system.  After
removing this scaling, we find that
for a system with extensive connectivity
and a bounded mass distribution,
the spectrum has a single zero mode and then a gap to the next eigenvalue
$E$.  For a system with intensive connectivity or with
unbounded masses, the gap is filled in by Griffiths effects.

To determine the localization properties of a normalized eigenfunction
$\psi_{i,l}$ associated with an eigenvalue $E_l$, we consider the
inverse participation ratio, $\sum_i |\psi_{i,l}|^4$.  In the
delocalized phase, this quantity scales with $1/N$.  
For constant mass\cite{local} and large connectivity, in the center of the band
the states are delocalized, while they are localized near the band edge.

We have found a very different result in the case of random masses.
With large connectivity and random masses, we find that all the 
states, with the exception of the zero mode, become localized as shown
in Fig.~1.  We plot the inverse participation ratio against eigenvalue,
for systems with bounded, random mass.
The solid line is $N=1000,p=1$ averaged
over $100$ samples while
the dashed is $N=500,p=1$ averaged over $1000$ samples.  
In the inset, we have divided the eigenvalue
by $N$, showing a perfect collapse of the two curves, indicating
that the states are localized.
\begin{figure}
\center{
\epsfxsize=2.5in
\epsfbox{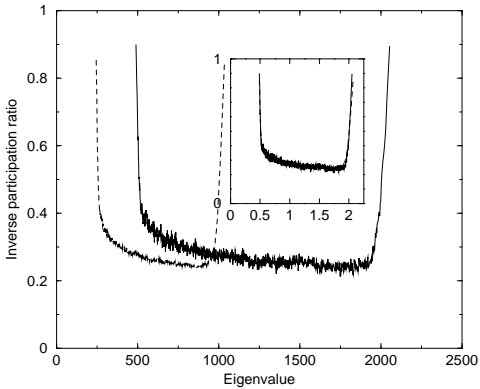}}
\vspace*{0.2cm}
\caption{Inverse participation ratio as a function of eigenvalue.  Solid
line is $N=1000,p=1$ with random, bounded mass.  Dashed line is same except
$N=500$.  Inset: eigenvalue has been scaled by system size.
}
\end{figure}

For extensive connectivity, localization holds even for very weak variation
in the masses.  This can be understood analytically via the renormalization.  
Let us
divide $L$ by $N$; then the $L_{ii}$ are all fluctuating variables of
order unity, with fluctuations, due to the random mass, which are also of
order unity.  The $L_{ij}$ are of order $1/N$.  Using the RG to remove
sites from the system at eigenvalue $E\neq 0$, removing a single site $j$
leads to corrections of order $L_{ij}^2/(L_{jj}-E)$.  The numerator is
of order $1/N^2$, while due to the random mass the denominator does not
vanish.  Thus, the dynamics of a single site is only weakly
affected by other sites, and the states are localized.

For low connectivity, we have considered systems with
100, 200, 400, 800, and 1600
sites, averaging over 1000 realizations for
systems with up to 800 sites, and over 100 realizations for the largest
systems, considering the three different mass distributions,
with average
connectivities of 1,2, and 4.
One expects that for low connectivity, greater than 1, the network
approximates an infinite dimensional system, or Bethe lattice.  Then,
at zero eigenvalue there is an extended state, and for low eigenvalues the 
variations in masses and spring constants only weakly scatter the vibrational 
waves.  Thus, one expects\cite{sl}
a delocalized phase for small eigenvalues for average connectivity
greater than $1$.

Numerically, the results for low connectivity are less clear.  For
average connectivity equal to $1$, the inverse participation ratio
is found to be independent of $N$, and thus the system appears localized.  For
average connectivity greater than $1$, the inverse participation ratio
decreases with increasing $N$; however, we do not observe a clear $1/N$ scaling
of the inverse participation ratio for an average connectivity of $2$.
For average connectivity $4$, we do see approximate $1/N$ scaling for
low $E$, indicating
delocalized states.

{\it Comparison of Eigenspectra---}
We have then tested the renormalization procedure against exact results
by applying the procedure to {\it single} realizations of the system.
We have found that in the correct regimes, the procedures are highly
accurate even for the details of specific realizations.  The procedure
work best for high connectivity, and for wide distributions of the masses,
with the advantage of the two-site technique being that it drives the
system to the regime in which it works accurately.
\begin{figure}
\center{
\epsfxsize=3in
\epsfbox{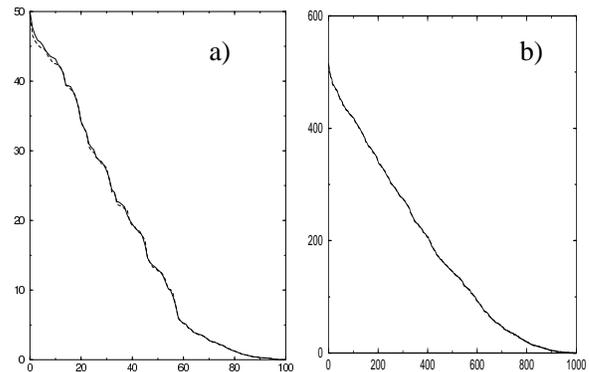}}
\vspace*{0.2cm}
\caption{a): $N=100, p=1$.  Exact result (solid) versus renormalization 
(dashed).
b): $N=1000, p=1$.  Exact result (solid) versus renormalization (dashed).
The distinction between the curves is not visible on the figure.
}
\end{figure}

In Fig.~2a, we show the results of the single-site technique for
a system with 100 sites, $p=1$, and the unbounded distribution of masses.
The sorted eigenvalues are plotted, with the $n$-th eigenvalue $E_n$ plotted
at position $(n,E_n)$.
The dashed line gives the results of the 
renormalization procedure, while the solid line is the exact result.  For
low eigenvalues, the two are indistinguishable.  In Fig.~2b, we show the
same for a system with 1000 sites; the two lines cannot be distinguished on
the figure.

At lower connectivity, with less randomness in the mass, the two-site
procedure becomes necessary.  In Fig.~3a, we show a system with $N=100, p=.01,$
and the bounded, random distribution of masses.  The solid line is the
exact result, the upper dashed line is the two-site result, and the lower
is the single-site result.

As one measure of the accuracy of the procedure, 
we sort the eigenvalues from highest to lowest,
and divide the root-mean-squared error in the $n$-th eigenvalue by the 
root-mean-squared value of that eigenvalue.  The result is quite
accurate, with the relative error averaging,
for example, $\lesssim 1\%$ for $N=100,p=1$ and bounded, random mass, becoming
even more accurate for larger systems.
As a more stringent test of the accuracy, if we instead divide the
root-mean-square error by the
root-mean-square {\it sample-to-sample}
fluctuations in that eigenvalue,
the relative error averaged over all $n$ is
 $\sim 15\%$ for the same system.
For lower connectivity, the relative error (compared to sample-to-sample
fluctuations)
is worse, $\sim 28\%$ for $N=100,p=.01$.  However, 
for the smaller eigenvalues, 
the accuracy increases:
for $N=100,p=.01$, the error relative to fluctuations
in the 80 lowest eigenvalues averages
$\sim 11 \%$.  

\begin{figure}
\center{
\epsfxsize=3in
\epsfbox{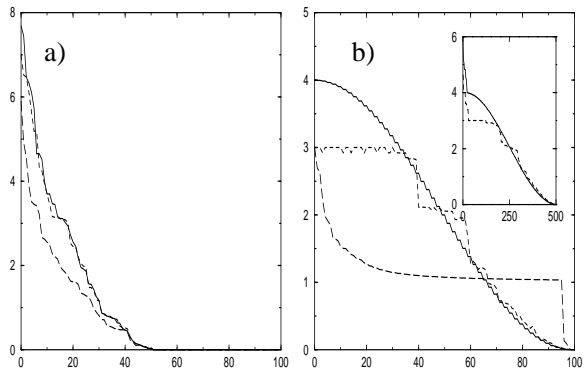}}
\vspace*{0.2cm}
\caption{a): $N=100, p=.01$.  Exact result (solid) versus two-site renormalization (upper dashed) and single-site renormalization (lower dashed). b):
$100$ site periodic chain.  Exact (solid), random choice of
sites for renormalization (dashed), poor choice (long-dashed).  Inset:
a small-world network.  Exact (solid), renormalization (dashed).
}
\end{figure}

Finally, we test the renormalization on a periodic 
one dimensional chain with constant masses and springs constants,
combining both low connectivity and no randomness.  If, in the two-site
procedure, when faced with a choice between different
sites with the same $L_{ii}$, we choose randomly, the low energy
properties are reasonable, while if we deliberately make poor choices, 
keeping the connectivity low, the results are much worse, as
shown in Fig.~3b.
Using a network improves on the results obtained\cite{dmrg} on a
one dimensional chain with a similar renormalization procedure, {\it
which preserves the one dimensional structure}.  
In the inset to Fig.~3b we consider a small-world network.  We take a
periodic, one dimensional chain with $500$ sites and constant mass, and
with probability $p=0.0002$ we connect pairs of sites which are not
already connected by the chain.

{\it Conclusion---}
In conclusion, we have introduced a real-space renormalization procedure
for studying certain linear operators on networks.  The procedure is exact
at low energy, and in many cases leads to very good results for the spectra
even at much higher energy.  This procedure can be used to obtain analytic
results on the localization properties.  It can also be used to obtain results 
for dynamics on a specific graph more rapidly than would be possible
with matrix diagonalization routines.  While for a fully connected
graph it still requires $O(N^3)$ time, the prefactor is much smaller
than that required for matrix diagonalization.  For less connected
networks, the procedure runs much more rapidly.  This difference will be 
important when studying large social networks which may have billions of nodes.

We have also found that the introduction of random masses leads to many
more localized states than constant masses.  For amorphous
systems, the random mass case is the relevant case.

The procedure naturally helps simplify the network by finding a simpler
network with a similar spectrum.  
Although we have considered only
linear problems, we hope that this kind of real-space procedure will
be an important tool in studying {\it nonlinear} problems in future
work, such as spin-glass or optimization problems.  

{\it Acknowledgments---}
I thank Z. Toroczkai for inspiration and discussions, and C. Jarzynski
for suggesting the relative error measurement.
This work was supported by DOE W-7405-ENG-36.

\end{document}